\documentclass[twocolumn,preprintnumbers,amsmath,amssymb,superscriptaddress,footnote]{revtex4-2}
\usepackage{graphicx} 
\usepackage[dvipdfmx]{epsfig} 
\usepackage{bm}
\usepackage{ulem}
\usepackage{amsmath,color}

\def\m@thcombine#1#2{
  \setbox0=\hbox{$#1$}
  \setbox1=\hbox{$#2$}
  \ifdim\wd0>\wd1
    \setbox0=\hbox to\wd1{\hss\box0\hss}
  \else
    \setbox1=\hbox to\wd0{\hss\box1\hss}
  \fi
  \mathop{\vcenter{
    \offinterlineskip\box0\box1}}}
\def\lesim{\m@thcombine<\sim}
\def\gesim{\m@thcombine>\sim}

\begin{document}

\title{Deposition of $\alpha$ clusters on the surface of $^{28}$Si}

\author{W. Horiuchi}
 \email{whoriuchi@omu.ac.jp}
\affiliation{Department of Physics, Osaka Metropolitan University, Osaka 558-8585, Japan}
\affiliation{Nambu Yoichiro Institute of Theoretical and Experimental Physics (NITEP), Osaka Metropolitan University, Osaka 558-8585, Japan}
\affiliation{RIKEN Nishina Center, Wako 351-0198, Japan}

\author{N. Itagaki}
 \email{itagaki@omu.ac.jp}
\affiliation{Department of Physics, Osaka Metropolitan University, Osaka 558-8585, Japan}
\affiliation{Nambu Yoichiro Institute of Theoretical and Experimental Physics (NITEP), Osaka Metropolitan University, Osaka 558-8585, Japan}
\affiliation{RIKEN Nishina Center, Wako 351-0198, Japan}

\author{Y. Yamaguchi}
\affiliation{Department of Physics, Osaka Metropolitan University, Osaka 558-8585, Japan}

\begin{abstract}
The possibility of the appearance of the clustering structure 
in the surface 
is one of the key features determining the nuclear 
structural properties. 
In this work, we study the ground-state structure of $^{28}\text{Si}$ using the antisymmetrized quasicluster model (AQCM), focusing on the deposition of the $\alpha$ cluster structure on the surface.
Three candidate structural configurations constrained to reproduce the experimental charge radius are investigated based on the $\text{SU}(3)$ shell model, the $j$-$j$ coupling shell model, and a localized $7\alpha$ cluster model. 
Using Glauber-model calculations for intermediate-energy proton elastic scattering, we demonstrate that the angular distribution at and beyond the first diffraction peak is highly sensitive to surface diffuseness, and that the $7\alpha$ cluster model configuration best reproduces the experimental data due to its sharper surface profile. 
Also, the elastic charge form factors from electron scattering support the existence of the cluster-induced oblately deformed structure.
\end{abstract}

\maketitle
  
\section{Introduction}

Atomic nuclei exhibit a rich diversity of structural 
phenomena resulting from the delicate interplay between 
independent-particle motion and many-body 
correlations~\cite{BohrMottelson1969}. 
While the traditional nuclear shell model successfully 
describes a vast array of nuclear properties in terms of 
single-particle motion within a mean 
field~\cite{Mayer1949,Haxel1949}, 
and picture dominates the nuclear structure
with increasing mass number owing to the spin-orbit interaction,
cluster structures 
frequently emerge in light and medium-mass systems as 
energetically favored, highly correlated 
configurations~\cite{Ikeda1968,Horiuchi1972,Freer2007,Oertzen2006}. 

The $\alpha$ decay has been known
as  an evidence 
that cluster correlation is important
in heavy nuclei~\cite{Gamow1928ZPhys,
Gurney1928Nature,Lovas1998PhR,Qi2019PPNP}.
Also, recent studies
have provided 
further important insights into the interplay between nuclear 
structure and radioactive decay dynamics.
The 
analysis of $\alpha$-decay properties serves as a sensitive 
probe of the underlying single-particle structure and 
pairing correlations, offering valuable information on the 
location of predicted shell closures and the possible 
existence of the island of stability~\cite{Typel2010PRC,Typel2014PRC,Typel2020EPJST,Tanaka2021Science}. 
Their theoretical 
approaches, often incorporating microscopic nuclear models 
and improved descriptions of nuclear matter properties, have 
also helped bridge the gap between experimental observations 
and fundamental theories of nuclear forces, providing a 
framework for predicting unknown heavy nuclei and guiding 
future experiments on the synthesis and characterization of 
superheavy elements.

For instance, the recent experimental studies of $\alpha$ decay in 
Te isotopes, which lie in a transitional region 
between spherical shell-model configurations and collective or 
cluster-like structures, offer valuable information 
on how $\alpha$-cluster components emerge in complex nuclear systems~\cite{Auranen2018PRL,Te104Nature2026}. The measured decay energies, half-lives, and hindrance factors 
serve as sensitive probes of $\alpha$ preformation probabilities and 
nuclear wave functions, allowing stringent tests of modern nuclear 
models. In particular, these studies contribute to a deeper 
understanding of the microscopic origin of the $\alpha$ decay and the 
evolution of nuclear structure with neutron–proton asymmetry, 
providing important benchmarks for theoretical approaches ranging 
from mean-field models to cluster models. The results 
therefore represent a significant step toward a unified description 
of radioactive decay phenomena and the structural evolution of 
exotic nuclei.

To account for the $\alpha$ decays, four nucleons (two protons 
and two neutrons) must be correlated around the nuclear surface
and form an $\alpha$ cluster.
The emergence of such surface-localized $\alpha$ clustering 
drastically alters the density profile, particularly the 
diffuseness around the nuclear surface. Recently, the 
present authors developed an efficient method to visualize 
ground-state nuclear structures by utilizing proton-nucleus 
elastic scattering combined with the antisymmetrized 
quasicluster model 
(AQCM)~\cite{Horiuchi22c,Horiuchi23,Yamaguchi2023,Okada2024}. 
We demonstrated that distinct shell and cluster 
configurations can be clearly differentiated by analyzing 
the angular distributions of the differential cross 
sections, particularly around the first diffraction peak.

The coexistence and competition between shell-model and 
cluster degrees of freedom have been widely debated in $p$-shell 
and light $sd$-shell nuclei, such as $^{12}$C, 
$^{16}$O, $^{20}$Ne, and $^{28}$Si. Among these, $^{28}$Si 
is of particular interest; its ground state exhibits a 
strong oblate deformation, yet it simultaneously 
accommodates cluster-like configurations featuring highly 
developed surface 
structures~\cite{Taniguchi2009,KanadaEnyo2011,Chiba2017,Ichikawa2011,Bauhoff1981,Hatakeyama19}.
Although the nucleus corresponds to the 
subclosure or $d_{5/2}$ in terms of the shell model, 
where the attractive
effect of the spin-orbit interaction is quite strong,
cluster correlation may be persistent and contributes 
for the oblate deformation.
In this work, we focus on $^{28}$Si as a prime candidate for 
hosting such a dual nature. We investigate the spatial 
distribution of the $\alpha$ clusters near the nuclear 
surface. 

To systematically analyze these effects, we construct three distinct structural configurations for $^{28}$Si: (a) the SU(3) shell model state, (b) the $j$-$j$ coupling shell model state, and (c) the $\alpha$ cluster model state. 
The charge radius of each configuration is constrained to reproduce the experimental value to ensure a meaningful comparison. 
Crucially, all three configurations are generated within the unified framework of the AQCM. 
This model enables a continuous, variationally controlled transition from a pure cluster structure to a $j$-$j$ coupling shell model state by systematically tuning the control parameters embedded in the many-body wave function~\cite{Itagaki2005,Itagaki2011,Suhara2013,Itagaki2016,Matsuno2017,Itagaki2019,Itagaki2020}. 
This framework has proven highly successful in investigating cluster-shell competition across various nuclear charts.

In the present study, we apply the AQCM wave functions of $^{28}$Si to Glauber-model calculations to assess how these intrinsic cluster structures manifest in proton elastic scattering observables. 
Cluster development near the surface directly influences the matter density's tail and surface diffuseness. 
Because intermediate-energy proton elastic scattering is highly sensitive to the spatial profile of the nuclear density, it serves as an ideal experimental probe. Specifically, we show how the resulting diffraction patterns and angular cross sections reflect the subtle changes in both the effective nuclear radius and the central density depletion induced by clustering.

This paper is organized as follows. Section~\ref{sec:method} describes the theoretical framework, detailing the preparation of the configurations using the AQCM and the formulation of proton and electron scattering. 
In Section~\ref{sec:results}, we analyze the potential emergence of clustering in the surface of $^{28}$Si and propose realistic experimental indicators. 
Finally, a summary and concluding remarks are provided in Section~\ref{sec:summary}.

\section{Method}
\label{sec:method}

\subsection{Configurations of $^{28}$Si within the AQCM}

The structural configurations of $^{28}$Si are generated using the antisymmetrized quasicluster model (AQCM)~\cite{Itagaki2005,Itagaki2011,Suhara2013,Itagaki2016,Matsuno2017,Itagaki2019,Itagaki2020}.
The AQCM is an extension of the conventional multi-$\alpha$ cluster model~\cite{Brink}, which provides a unified framework capable of expressing not only localized cluster states but also the $j$-$j$ coupling and SU(3) shell-model configurations within a single scheme. 
The AQCM basis state for $^{28}$Si is formulated as a $7\alpha$ cluster system:
\begin{align}
  \Phi(\nu,d,\Lambda)=\mathcal{A}
  \left\{\prod_{i=1}^7 \Phi_\alpha(\nu,\bm{R}_i,\Lambda)\right\},
\end{align}
where $\mathcal{A}$ denotes the antisymmetrizer for all constituent nucleons. 
Here, $\Phi_{\alpha}(\nu,\bm{R}_i,\Lambda)$ represents the wave function of the $i$-th $\alpha$ particle, with its Gaussian center parameter $\bm{R}_i$ defined via single-nucleon states. It is written as
\begin{align}
\Phi_{\alpha}(\nu,\bm{R}_i,\Lambda)=
\phi^{\nu \bm{R}_i\Lambda}_{1\uparrow p}\phi^{\nu \bm{R}_i\Lambda}_{2\downarrow p}
\phi^{\nu \bm{R}_i\Lambda}_{3\uparrow n}\phi^{\nu \bm{R}_i\Lambda}_{4\downarrow n},
\end{align}
where the single-nucleon Gaussian wave packet, including the spin $\chi_s$ ($s=\uparrow$ or $\downarrow$) and isospin $\eta_t$ ($t=p$ or $n$) wave functions, is given by
\begin{align}
\phi_{jst}^{\nu \bm{R}_i\Lambda}=  \left(\frac{2\nu}{\pi}\right)^{3/4}\exp\left[-\nu(\bm{r}_j-\bm{\zeta}(\bm{R}_i,\Lambda))^2\right]\chi_{s}\eta_{t}.
  \label{spwf}
\end{align}
The complex parameter $\bm{\zeta}$ incorporates the spin-orbit correlation and is defined as
\begin{align}
  \bm{\zeta}(\bm{R}_i,\Lambda)=\bm{R}_i+i\Lambda \bm{e}^{\rm spin}\times \bm{R}_i,
\end{align}
where $\bm{e}^{\rm spin}$ is the unit vector representing the intrinsic-spin orientation of the nucleon.

The intrinsic structure of $^{28}$Si is described by following the insight of microscopic antisymmetrized molecular dynamics calculations, in which a pentagonal $7\alpha$-like structure emerges in the oblate intrinsic state of $^{28}$Si without assuming $\alpha$ clustering {\it a priori}~\cite{KanadaEnyo2011}. More specifically, the
intrinsic basis function of $^{28}$Si is constructed from five $\alpha$ clusters arranged in a pentagonal configuration within the $xy$-plane, supplemented by two additional $\alpha$ clusters positioned perpendicularly along the $z$-axis. 
For the pentagonal component in the $xy$-plane, each $\alpha$ particle is placed at a common distance $d$ from the origin. 
The remaining two $\alpha$ clusters are located at $d/2$ and $-d/2$ along the $z$-axis.
Notably, one can recover the SU(3) shell-model configuration, $[(0s)^4(0p)^{12}(1s0d)^{12}]$, by taking the limit $d \to 0$~\cite{Brink} with $\Lambda=0$. 
The choice of this pentagon configuration on the $xy$-plane is almost
inevitable from the viewpoint of the SU(3) limit. 
Other possible geometrical arrangements do not lead to
the lowest principal quantum number configuration in the limit $d \to 0$.
For example, a square configuration on the $xy$-plane combined with
three $\alpha$ clusters aligned along the $z$-axis, or a hexagon
configuration on the $xy$-plane with the remaining $\alpha$ cluster
located at the center, cannot evolve into the lowest SU(3)
state. Conversely, with this specific geometric arrangement of $\alpha$ particles, taking the limit $d \to 0$ while setting $\Lambda=1$ yields the $j$-$j$ coupling shell-model state, which corresponds to the $d_{5/2}$ subclosure configuration, $[(0s)^4(0p)^{12}(0d_{5/2})^{12}]$.
We note, however, that this specific distance of the two $\alpha$ clusters from the pentagonal plane is not microscopically constrained in the present model. In general, moving the $\alpha$ clusters outward makes the surface density more diffuse and enhances the peak of the elastic scattering cross section, whereas moving them inward has the opposite effect. In a realistic wave function, such different configurations should be superposed and collectively contribute to the surface diffuseness of $^{28}$Si. However, obtaining such a realistic wave function while simultaneously reproducing the experimental nuclear radius, which is essential for the present discussion of scattering cross sections, is a challenging task and is beyond the scope of the present study. Among the configurations considered above, as will be shown later, the present choice is found to provide the best description of the first peak of the elastic scattering cross section and is therefore adopted in the present calculation.

For each AQCM wave function, the nucleon one-body density distribution in the body-fixed frame for a nucleus with mass number $A$ is evaluated as
\begin{align}
  \tilde{\rho}(\bm{r})
  = \langle \Phi |\sum_{i=1}^A \delta(\bm{r}_i-\bm{r}) | \Phi \rangle / \langle \Phi | \Phi \rangle,
\end{align}
where $\sum_{i=1}^A\left<\Phi\right|\bm{r}_i\left|\Phi\right>=0$. 
Because $\tilde{\rho}(\bm{r})$ generally includes spurious center-of-mass motion, this component is properly removed following the prescription described in Ref.~\cite{Horiuchi07}:
\begin{align}
  \int d\bm{r}\,e^{i\bm{k}\cdot\bm{r}}\rho_{\rm int}(\bm{r})=\exp\left(\frac{k^2}{8A\nu}\right)
  \int d\bm{r}\,e^{i\bm{k}\cdot\bm{r}} \tilde{\rho}(\bm{r}).
\end{align}
The density distribution in the laboratory frame is subsequently obtained by taking the angular average of the center-of-mass-free, body-fixed density distribution~\cite{Horiuchi12}:
\begin{align}
  \rho(r)=\frac{1}{4\pi}\,\int d\hat{\bm{r}}\,\rho_{\rm int}(\bm{r}).
\label{density.eq}
\end{align}

\subsection{Elastic scattering cross sections}

To establish a link between the calculated density profiles and empirical reaction observables, we compute the elastic scattering cross sections using the Glauber model~\cite{Glauber}, a well-established high-energy microscopic reaction theory. Below, we briefly outline the procedure for deriving these cross sections from the calculated one-body density distributions.

The scattering amplitude for proton-nucleus elastic scattering is expressed as~\cite{Suzuki03}
\begin{align}
  f(\theta)=F_C(\theta)+\frac{ik}{2\pi}\int d\bm{b}\, e^{-i\bm{q}\cdot\bm{b}+2i\eta \ln(kb)}\left(1-e^{i\chi_{pT}(\bm{b})}\right),
\end{align}
where $F_C(\theta)$ is the Rutherford scattering amplitude, $\bm{b}$ is the impact parameter vector, and $\eta$ is the Sommerfeld parameter. 
Relativistic kinematics
is
adopted for the wave number $k$. 
Using this scattering amplitude, the differential cross section for proton-nucleus elastic scattering is given by
\begin{align}
  \frac{d\sigma}{d\Omega}=|f(\theta)|^2.
\end{align}

The optical phase-shift function $\chi_{pT}$ evaluated within the optical-limit approximation (OLA) takes the form~\cite{Glauber,Suzuki03}
\begin{align}
  i\chi_{pT}(\bm{b})\approx -\int d\bm{r}\,
  \left[\rho_p(\bm{r})\Gamma_{pp}(\bm{b}+\bm{s})
  +\rho_n(\bm{r})\Gamma_{np}(\bm{b}+\bm{s})\right],
\end{align}
where $\rho_p=\rho_n=(1/2)\rho$ 
and $\bm{r}=(\bm{s},z)$, with $z$ aligned along the beam direction. 
The validity of this approximation
has been demonstrated in Refs.~\cite{Varga02,IbrahimJPSJ,Nagahisa18,Hatakeyama19} and further established in Refs.~\cite{Horiuchi2026a,Horiuchi2026b}.
The primary inputs for this framework are the density distributions obtained from Eq.~(\ref{density.eq}) and the elementary proton-proton (neutron-proton) profile functions $\Gamma_{pp}$ ($\Gamma_{np}$). 
We adopt the profile function parameterizations from Ref.~\cite{Ibrahim08}, which are well tested and have demonstrated reliable performance in proton-nucleus scattering applications~\cite{IbrahimJPSJ,Horiuchi16,Hatakeyama19,Horiuchi22c,Horiuchi23}.
As discussed in Ref.~\cite{Hatakeyama18}, proton elastic scattering is highly sensitive to the density profile near the nuclear surface, rendering it a powerful spectroscopic tool for distinguishing between shell and cluster configurations in light- and medium-mass nuclei~\cite{Horiuchi22c,Horiuchi23,Yamaguchi2023,Okada2024}.

\subsection{Elastic charge form factor}

The elastic charge form factor obtained via electron scattering has traditionally served as a direct probe of internal nuclear charge distributions~\cite{Hofstadter56}. 
Because the electromagnetic electron-nucleus interaction is weaker than the hadronic proton-nucleus interaction, the elastic charge form factor is more deeply sensitive to the interior features of the charge density distribution. 
The elastic charge form factor is evaluated from the Fourier transform of the density distribution, convolved with the finite size of the proton charge~\cite{Kamimura81,Horiuchi23}:
\begin{align}
  |F(q)|^2=\left|\frac{1}{A}\int_0^{\infty}dr\, \rho(r)j_0(qr) r^2\right|^2
  \exp\left(-\frac{1}{2}a_p^2q^2\right),
  \label{form.eq}
\end{align}
where $a_p^2=0.514$~fm$^2$ is chosen to reproduce the charge radius of a free proton ($0.878$~fm)~\cite{Angeli13}.

\section{Results}
\label{sec:results}

\begin{figure}[t]
\centering
\includegraphics[width=\linewidth]{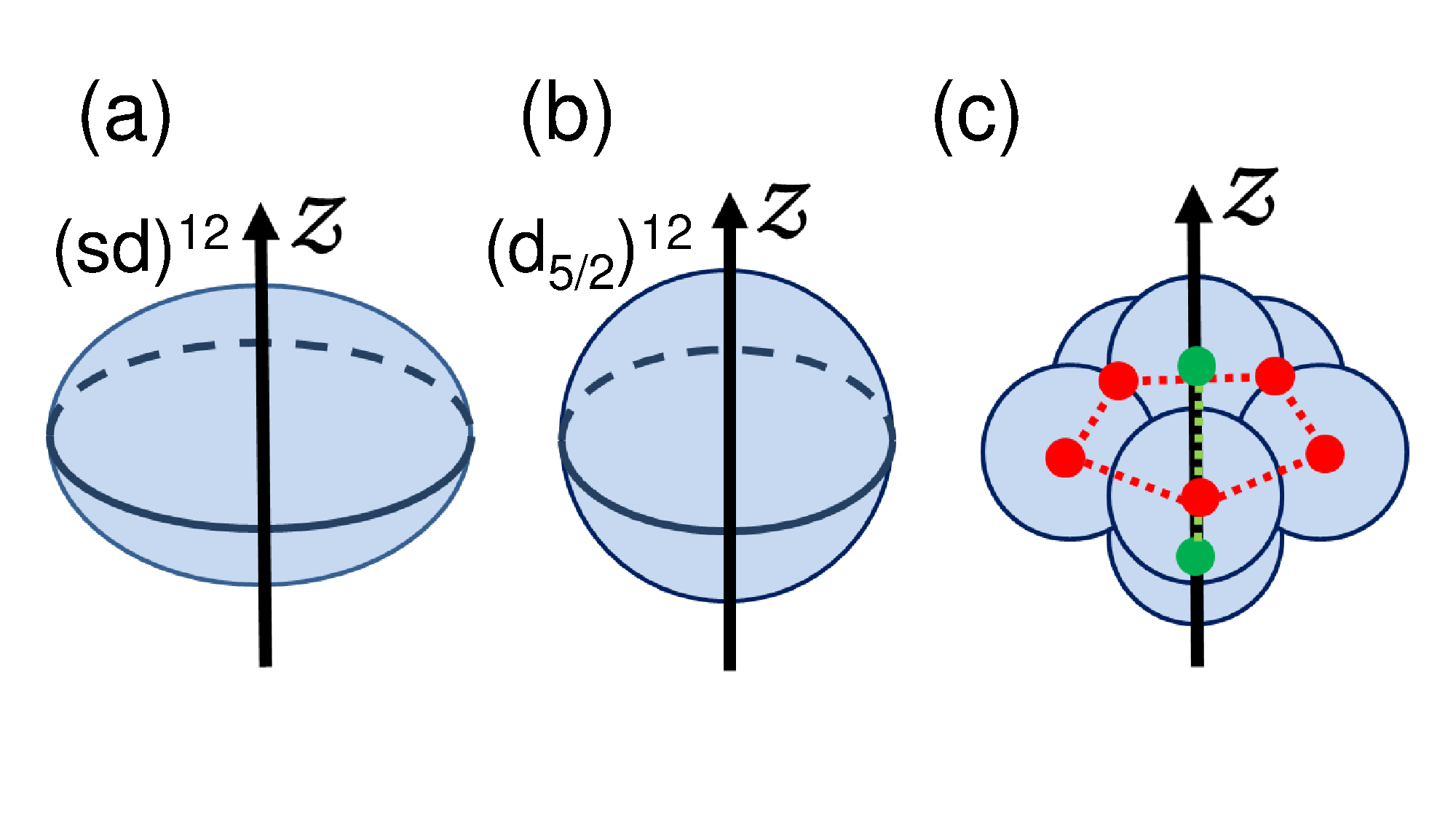}
\caption{Schematic illustration of the three structural configurations of $^{28}$Si generated using the AQCM: (a) SU(3) shell-model state, (b) $j$-$j$ coupling shell-model state, and (c) localized $7\alpha$ cluster state.}
\label{fig:schematic}
\end{figure}

We investigate three distinct structural configurations for $^{28}$Si constructed within the AQCM framework: the SU(3) shell-model limit, the $j$-$j$ coupling shell-model limit, and a localized cluster state, as illustrated schematically in Fig.~\ref{fig:schematic}. 
Here, the SU(3) state represents a seven-$\alpha$ configuration evaluated in the zero-distance limit ($d \to 0,$ $\Lambda=0$) of the relative motion between the clusters. The $j$-$j$ shell state ($d\to 0,$ $ \Lambda=1)$ corresponds to the ideal $d_{5/2}$ subclosure configuration $[(0s)^4(0p)^{12}(0d_{5/2})^{12}]$ of the traditional nuclear shell model. 
In contrast, the cluster configuration consists of seven $\alpha$ clusters 
with the sizes of ``free" ones held at finite relative distances, preserving a five-fold rotational symmetry in its intrinsic frame. 

All three configurations are constrained to precisely reproduce the experimentally observed ground-state charge radius of $3.1124(24)$~fm~\cite{Angeli13}. 
For the SU(3) and $j$-$j$ shell states, this constraint is satisfied by tuning the size parameter $\nu$ of the single-particle wave function in Eq.~(\ref{spwf}).
The resulting size parameter
is set to $\nu=0.1508$~fm$^{-2}$ for both configurations.
For the cluster state, the single-particle width parameter is fixed to that of a free $\alpha$ particle ($\nu = 0.2656$~fm$^{-2}$), while 
the relative distance between the clusters is set to $d=2.639$~fm to reproduce the ground-state charge radius of $^{28}$Si. 
We also calculate the quadrupole deformation parameter $\beta_2$, defined as
\begin{align}
\beta_2=\frac{4\pi}{5}\frac{\langle r^2Y_{20}\rangle}{\langle r^2\rangle},
\end{align}
where $\langle r^2Y_{20}\rangle$ and $\langle r^2\rangle$ denote the intrinsic quadrupole moment and the mean-square radius, respectively.
The calculated values are $\beta_2=-0.221$ for the SU(3) shell state, $\beta_2=0.018$ for the $j$-$j$ coupling shell state, and $\beta_2=-0.385$ for the cluster state. As expected, the SU(3) shell state exhibits an oblate deformation, while the $j$-$j$ coupling shell state is nearly spherical. The small but finite deformation arises from the use of $d=0.01$ fm instead of the numerical limit $d\to 0$ and has a negligible effect on the results.
The cluster state shows a more pronounced oblate deformation, which is consistent with the deformation inferred from the $B(E2)$ value~\cite{Pritychenko2016}.

\begin{figure}[t]
\centering
\includegraphics[width=\linewidth]{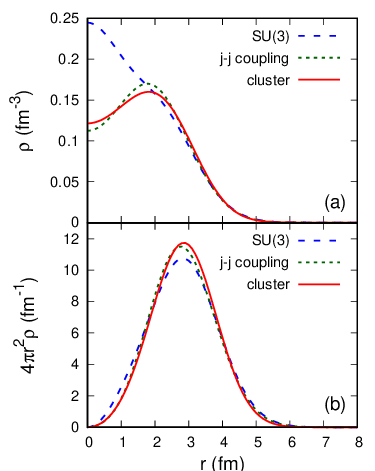}
\caption{Nucleon density distributions for the three AQCM configurations of $^{28}$Si: (a) $\rho(r)$ and (b) $4\pi r^2\rho(r)$}.
\label{fig:density}
\end{figure}

We first examine the nucleon density distributions in the laboratory frame derived from these three AQCM wave functions.
Figure~\ref{fig:density} displays the calculated radial nucleon density distributions. 
To better illustrate the density distributions in the inner and surface regions, we plot both $\rho(r)$ and $4\pi r^2\rho(r)$.
As expected, the SU(3) shell state shows an enhanced central density owing to the presence of the $1s$ orbit. In contrast, the $j$-$j$ coupling shell state exhibits a bubble structure because it represents a spherical sub-shell-closed configuration in which the $1s$ orbit is completely unoccupied.
The cluster configuration exhibits a markedly different distribution relative to the SU(3) and $j$-$j$ shell states. 
Crucially, $\alpha$-cluster formation enhances the matter distribution in the nuclear surface region while maintaining the identical charge radius. 

This structural divergence becomes apparent when extracting the surface diffuseness parameter, $a$. We characterize the surface profile of each configuration by fitting the calculated densities to a standard two-parameter Fermi (2pF) distribution as:
\begin{equation}
\rho_{\rm 2pF}(r; R, a) = \frac{\rho_0}{1 + \exp\left(\frac{r - R}{a}\right)},
\label{Fermi}
\end{equation}
where $\rho_0$ is the normalization constant, which is uniquely determined for given $R$ and $a$. Following Ref.~\cite{Hatakeyama18}, 
the fitting is performed by minimizing
\begin{align}
\int_0^\infty dr\, r^2|\rho_{\rm 2pF}(r; R,a)-\rho(r)|.
\end{align}
This procedure is well suited for reproducing the surface region of the theoretical density distribution.
The extracted diffuseness parameters 
are 
0.532~fm for the SU(3) shell state, 0.483~fm for the $j$-$j$ shell state, and 0.450~fm for the cluster state.
The diffuseness parameter of the SU(3) shell state is consistent with the typical
diffuseness parameter for stable nuclei about 0.54~fm~\cite{BohrMottelson1969}. For the $j$-$j$ shell state,
the nuclear surface becomes sharper owing to the
absence of $1s$ orbits~\cite{Choudhary2020}.
Notably, the cluster configuration yields 
the smallest diffuseness parameter of $0.450$~fm, which is roughly 10--20\% smaller than those of the shell-model configurations. 
Because the nucleons are localized into tightly bound $\alpha$ clusters residing near the surface, the nuclear diffuseness becomes extremely narrow and the central region of the nucleus experiences a severe matter deficit, naturally generating a central bubble structure.

Despite all configurations sharing a common charge radius that matches experimental data, their radial density profiles differ dramatically near both the center and the surface. 
This mismatch demonstrates that bulk properties, such as the charge radius, are insufficient to uniquely identify cluster correlations, highlighting the necessity of surface-sensitive reaction observables.

\begin{figure}[t]
\centering
\includegraphics[width=\linewidth]{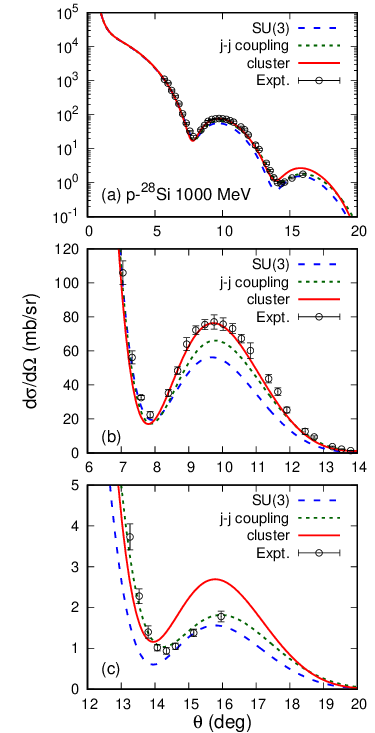}
\caption{Calculated proton elastic scattering differential cross sections at 1000 MeV/nucleon for the three AQCM configurations compared with experimental data. The panel (a) displays the cross sections on a logarithmic scale, while the panels (b) and (c) focus on the first and second diffraction peaks on a linear scale, respectively. The experimental data is taken from Ref.~\cite{Alkhazov1976}.}
\label{fig:p-scattering_three}
\end{figure}

Using the matter density distributions obtained from the AQCM, we evaluate proton-nucleus elastic scattering cross sections within the framework of the Glauber model. 
Figure~\ref{fig:p-scattering_three} presents the calculated differential cross sections as a function of the center-of-mass scattering angle. While the general diffraction pattern is reproduced on a logarithmic scale for all three states in Fig.~\ref{fig:p-scattering_three} (a), notable configuration-dependent differences emerge when viewed on a linear scale around the first and second diffraction peak in Fig.~\ref{fig:p-scattering_three} (b) and (c), respectively.
Intriguingly, the magnitude of the first diffraction peak at $\theta \approx 10^\circ$ scales inversely with the surface diffuseness of the configurations. 
The cluster configuration, characterized by its sharper surface density fall-off, exhibits the highest peak intensity and an enhanced cross section at larger angles. 
This behavior stems directly from the modification of the surface density profile under clusterization. 
While forward-angle scattering is predominantly dictated by the bulk nuclear size, the cross sections at and beyond the first diffraction peak exhibit an elevated sensitivity to the surface diffuseness~\cite{Hatakeyama18}. 
Consequently, the angular distribution past the first diffraction minimum serves as a potent spectroscopic signature to separate cluster structures from mean-field configurations. 
Notably, as clearly seen in Fig.~\ref{fig:p-scattering_three} (b), only the cluster configuration successfully reproduces the experimental scattering data, exhibiting the sharpest surface diffuseness among the three configurations employed in this work. Although the $j$-$j$-coupling shell-model configuration also exhibits a sharp nuclear surface associated with the absence of the $s$-orbital, it is still insufficient to reproduce the experimental data. In contrast, as highlighted in Fig.~\ref{fig:p-scattering_three} (c), the shell-model configurations provide the best agreement with the cross sections around the second peak. The differences among the models, however, are not large enough to discriminate between them. This reflects the limited sensitivity of proton scattering to the underlying nuclear configurations.

\begin{figure}[t]
\centering
\includegraphics[width=\linewidth]{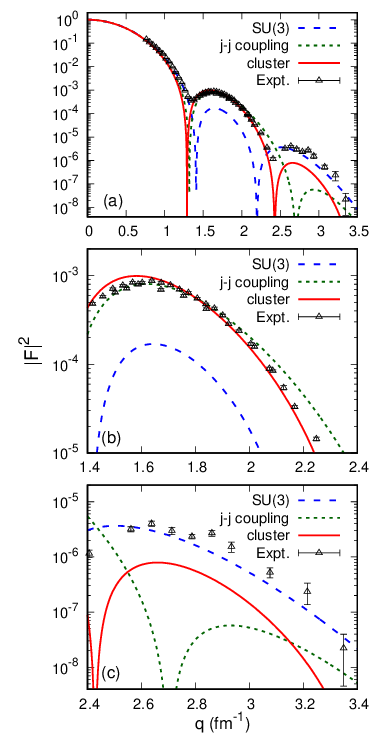}
\caption{Elastic charge form factors obtained by the three AQCM configurations compared with the experimental data. 
Panel (a) presents the full $q$ range to show the overall differences in the diffraction minima (dip positions), whereas panels (b) and (c) enlarge the regions $q=1.4$–2.4 fm$^{-1}$
and $q=2.4$-3.4 fm$^{-1}$, respectively, to emphasize the corresponding features.
The experimental charge form factor is extracted from Ref.~\cite{Li1974} with the Coulomb correction for $q$ prescribed in Ref.~\cite{Beck1984}.}
\label{fig:e-scattering}
\end{figure}

In Fig.~\ref{fig:e-scattering}, the calculated elastic charge form factors are plotted 
together with available data from electron scattering experiments as a function of the momentum transfer $q$.
In Fig.~\ref{fig:e-scattering} (a), since all configurations have the same charge radius, the results are identical at low $q$. As $q$ increases, however, differences among the configurations begin to emerge. Fig.~\ref{fig:e-scattering} (b) enlarges the region around the second peak, where the SU(3) result clearly deviates from the others. At even higher $q$, as shown in Fig.~\ref{fig:e-scattering} (c), the SU(3) result provides the best agreement with the experimental data.
This divergence is rooted in the intrinsic sensitivity of the electron probe. 
Unlike hadrons, electrons interact via the electric 
force and can penetrate the dense nuclear core without experiencing strong absorption. 
The form factor is the Fourier transform of the ground-state charge density distribution, as defined in Eq.~(\ref{form.eq}). 
High-momentum transfer regimes probe progressively shorter spatial distances, making the form factor at higher $q$ values deeply sensitive to the detailed variations of the interior density matrix. 

Indeed, at higher momentum transfer, the SU(3) shell state provides the best agreement with the experimental data. This result suggests that the interior of the nucleus is characterized by an SU(3)-like shell-model structure, whereas the surface region is dominated by a cluster-oriented oblate configuration. Such a coexistence of shell- and cluster-model features may reflect the radial dependence of the underlying nuclear structure. In this context, it would be interesting to discuss the role of the spin-orbit interaction, which is known to favor the $j$-$j$ shell-model configuration in the nuclear interior region. The present results imply that
although the proton and neutron numbers of 14 in $^{28}\mathrm{Si}$ 
correspond to the subclosure of the spin-orbit favored orbits, $d_{5/2}$,
SU(3)-like oblate deformation and $\alpha$ clustering in the surface are more important. 
Such competition of deformation/clustering and subclosure of spin-orbit favored orbits 
of the $j$-$j$ coupling shell model is discussed in $^{12}\mathrm{C}$,
which has the proton and neutron numbers of 6~\cite{Itagaki2004,Otsuka2022}.
Also, similar discussion has been extended to Ti region, where the nuclear surface is predominantly described by shell-model configurations, while cluster correlations emerge in the dilute tail region~\cite{Okada2024}.

\section{Summary and Perspective}
\label{sec:summary}

In summary, we have investigated the ground-state structural properties of $^{28}\mathrm{Si}$ within a unified framework using the antisymmetrized quasicluster model (AQCM) combined with high-energy microscopic reaction theory. 
To evaluate the competing behavior between mean-field single-particle degrees of freedom and spatial cluster correlations, we constructed three candidate structural configurations: the $\mathrm{SU}(3)$ shell-model state, the $j$-$j$ coupling shell-model state, and a localized $7\alpha$ cluster state. 
All three configurations were constrained to reproduce the experimental charge radius of $^{28}$Si, ensuring an unbiased basis for comparing their spatial distributions.

Our analysis of the radial nucleon density distributions reveals that the localized $7\alpha$ cluster configuration exhibits a highly modified surface profile compared to the shell-model limits. 
By parameterizing the calculated densities using a standard two-parameter Fermi distribution, we demonstrated that the cluster configuration is characterized by a significantly sharper surface diffuseness parameter, $a = 0.450~\mathrm{fm}$,
(cf. typical value $a=0.54$ fm~\cite{BohrMottelson1969}). 
This sharp surface fall-off results directly from the localization of nucleons into tightly bound, compact $\alpha$ clusters near the nuclear surface.

To evaluate the experimental visibility of these structures, we performed high-energy 
proton-nucleus scattering calculations by using the Glauber model.
The calculated cross sections demonstrated that the angular distribution at and beyond the first diffraction peak ($\theta \approx 10^\circ$) is highly sensitive to the surface diffuseness of the target nucleus. 
Among the evaluated models, only the $7\alpha$ cluster configuration successfully reproduces the absolute magnitude and diffraction pattern observed in experimental data, providing robust confirmation of localized alpha clustering at the nuclear surface. 

The results of both electron scattering and proton elastic scattering indicate that the cluster configuration provides the best reproduction of the experimental data around the first diffraction peak, suggesting the presence of cluster correlations in the nuclear surface region. In contrast, at the second diffraction peak, which is more sensitive to the inner part of the density distribution, the SU(3) shell-model oblate configuration yields the best agreement with the experimental data. These findings suggest that the intrinsic structure of the nucleus undergoes a gradual transition from an SU(3)-like shell-model configuration in the interior to a cluster-like configuration toward the surface. Although the present discussion is based on ideal intrinsic configurations, the results strongly motivate the development of a fully microscopic description capable of describing such a radial evolution of nuclear structure within a unified framework.

\section*{acknowledgments}

This work was in part supported by JSPS KAKENHI Grants
Nos.\ 23K22485, 25K01005, 25K07285, 26K00703, and JSPS Bilateral Program No. JPJSBP120247715.

\bibliography{si28bib.bib}

\end{document}